\documentstyle[epsf,eqsecnum,floats,preprint,aps]{revtex}

\tighten

\input epsf
\begin{document}

\newcommand{\be}{\begin{equation}}
\newcommand{\ee}{\end{equation}}
\newcommand{\bea}{\begin{eqnarray}}
\newcommand{\eea}{\end{eqnarray}}
\newcommand{\PSbox}[3]{\mbox{\rule{0in}{#3}\includegraphics{#1}\hspace{#2}}}

\def\5M{M^3_{(5)}}
\def\4M{M^2_{(4)}}

\overfullrule=0pt
\def\Int{\int_{r_H}^\infty}
\def\d{\partial}
\def\e{\epsilon}
\def\M{{\cal M}}
\def\high{\vphantom{\Biggl(}\displaystyle}
\catcode`@=11
\def\@versim#1#2{\lower\p@\vbox{\baselineskip\z@skip\lineskip-.5\p@
    \ialign{$\m@th#1\hfil##\hfil$\crcr#2\crcr\sim\crcr}}}
\def\simge{\mathrel{\mathpalette\@versim>}} %
\def\simle{\mathrel{\mathpalette\@versim<}} %
\def\sun{\hbox{$\odot$}}
\catcode`@=12 

\rightline{CERN--TH/2003--234}
\rightline{astro-ph/0310005}
\vskip 1cm

\setcounter{footnote}{0}

\begin{center}
\large{\bf Squeezing MOND into a Cosmological Scenario}
\ \\
\ \\
\normalsize{Arthur Lue\footnote{E-mail: lue@cern.ch}
  and Glenn D. Starkman\footnote{E-mail: glenn.starkman@cern.ch}}
\ \\
\ \\
\small{\em Center for Education and Research in Cosmology
  and Astrophysics\\
Department of Physics\\
Case Western Reserve University\\
Cleveland, OH 44106--7079\\
\ \\
and\\
\ \\
CERN Theory Division\\
CH--1211 Geneva 23\\
Switzerland}
\ \\

\end{center}

\begin{abstract}

\noindent
Explaining the effects of dark matter using modified gravitational
dynamics (MOND) has for decades been both an intriguing and
controversial possibility.  By insisting that the gravitational
interaction that accounts for the Newtonian force also drives cosmic
expansion, one may kinematically identify which cosmologies are
compatible with MOND, without explicit reference to the underlying
theory so long as the theory obeys Birkhoff's law.  Using this
technique, we are able to self-consistently compute a number of
quantities of cosmological interest.  We find that the critical
acceleration $a_0$ must have a slight source-mass dependence
($a_0\sim M^{1/3}$) and that MOND cosmologies are naturally
compatible with observed late-time expansion history and the
contemporary cosmic acceleration. However, cosmologies that can
produce enough density perturbations to account for structure
formation are contrived and fine-tuned. Even then, they may be
marginally ruled out by evidence of early ($z \sim 20$) reionization.
\end{abstract}

\setcounter{page}{0}
\thispagestyle{empty}
\maketitle

\eject

\vfill

\baselineskip 18pt plus 2pt minus 2pt

\section{Introduction}

That approximately ninety percent of the matter in the universe is
composed of some as of yet unspecified material is an unsettling
prospect, especially as an increasingly coherent picture of cosmology
emerges.  And while there are several 
well-motivated candidates for dark matter, its sole function
is to provide gravitational ballast by offering supplemental mass in
astrophysical and cosmological settings where the accounting of
visible matter falls short of gravitational expectations or requirements.
Replacing dark matter with a modification
of the laws of gravity (as encoded in the paradigm of Modified
Newtonian Dynamics, MOND) has been for decades both an
intriguing and a controversial alternative \cite{MOND,Sanders:2002pf}.

Nevertheless, the dark matter paradigm works quite well and
offers a litany of successes in its contribution to the standard
cosmological model.  In contrast, MOND, though not totally silent,
is largely inarticulate concerning cosmology:  it is a paradigm
designed to address galaxy rotation curves, there exists no
satisfactory underlying theory, and there is some difficulty in
incorporating it into a believable cosmological scenario 
\cite{Felten1984,McGaugh:1998vb,Sanders2001,Nusser:2001fx}.

In this paper, we reexamine the possibility of folding MOND into
a cosmological model under the premise that the same gravitational
interactions that manifest themselves in a modified Newtonian
force are also responsible for cosmological evolution.  In a previous
paper \cite{Lue:2003ky} with Scoccimarro, we devised a technique
by which one may kinematically derive a unique Schwarzschild-like
metric for a modified gravity theory from a specified, nonstandard
homogeneous cosmology.  Again, the presumption exploited was
that cosmology is driven exclusively by gravitational
self-interactions of the constituent matter, rather than by
some unknown energy-momentum component such as dark energy.  This correspondence between the metric and cosmology
can be made completely without reference to the fundamental
modified-gravity theory, using only the assumption that the underlying
theory respects Birkhoff's law.  The procedure is simply the
generalization of the classic description of how one recovers the
Friedmann equation from
the Newtonian force law, but generalized to a full metric theory and
beyond Einstein gravity.

We apply the same technique here to ascertain which cosmologies
are compatible with MOND, allowing us to identify a full
Schwarzschild-like metric and providing a self-consistent framework
to perform calculations of interest in MOND cosmology.  We begin by
briefly reviewing the prescription for the full metric consistent with
homogeneous cosmologies and apply that prescription to determine both
the Schwarzschild-like metric and the modified Friedmann equation of MOND.
Then we examine the class of cosmologies consistent with the MOND
force law and reveal that there are potentially insurmountable
difficulties that arise when one wishes to incorporate MOND into a
consistent cosmological scenario.

\section{Metric MOND and Local Gravity}

Let us quickly review the technique developed in
Ref.~\cite{Lue:2003ky} where one infers the Schwarzschild metric from
an arbitrary cosmology, to see how one might apply it in reverse and
devise a cosmology consistent with Modified Newtonian Dynamics.
Consider a homogeneous cosmology described by the line element
\be
	ds^2 = dt^2 - a^2(t)\delta_{ij}dx^idx^j\ ,
\label{line-cosmo}
\ee
with some specified scale factor evolution, $a(t)$.  If this universe
is matter-dominated (such that the Universe is filled homogeneously by
matter whose density obeys the relationship $\rho(t) \sim a^{-3}$),
one is faced with either of two possibilities.  First, one may believe
that Einstein gravity is correct and that the cosmology is driven by
some unseen additional energy-momentum components -- dark matter and dark energy.
The other possibility
is that the matter we see is the only energy-momentum component and
that one needs to alter gravitational dynamics in a specific way to
achieve the observed cosmic expansion history, $a(t)$.  We follow the
latter possibility.

It is convenient to represent the given scale factor evolution as the
solution to some alternative Friedmann equation:
\be
	H^2 \equiv {\dot{a}^2\over a^2} = H_0^2g(x)\ ,
\label{FRW}
\ee
where $x = {8\over 3}\pi G\rho/H_0^2$ is a dimensionless parameter,
$G$ is Newton's constant, and $H_0$ is today's Hubble scale. The
function $g(x)$ is determined by the given $a(t)$.  If one requires
that the fundamental gravitational theory respects Birkhoff's law, then
one can uniquely determine the metric of a spherically symmetric
source \cite{Lue:2003ky}.  That metric is described by the line
element
\be
	ds^2 = g_{00}(r)dt^2 - g_{rr}(r)dr^2 - r^2d\Omega\ ,
\ee
with
\be
	g_{00}(r) = g^{-1}_{rr} = 1 - r^2H_0^2g\left(r_g/r^3H_0^2\right)\ .
\label{metric}
\ee
Here $r_g = 2GM$, is the usual Schwarzschild radius of a matter
source of mass $M$.  Note that the form of the metric components
is completely determined by $a(t)$, and in particular, that
$r_g$ and $r$ can only appear in the metric in a specific combination.
This point will be important when we consider how to apply this
connection between cosmology and the Schwarzschild-like metric to
MOND.

In MOND\cite{MOND} the gravitational
acceleration exerted by a body of mass $M$ obeys the 
relationship:
\be
	a = \cases{ -{1\over 2}{dg_{00}\over dr} = -GMr^{-2}  & $\vert a\vert > a_0$ \cr
	\sim  -r^{-1} & $\vert a\vert < a_0$ \cr } \label{preMOND}
\ee
for some critical acceleration, $a_0$.  If we insist on a form for
modified gravity which is compatible with a homogeneous cosmology and
Birkhoff's law, its Schwarzschild-like metric must be of the form
Eq.~(\ref{metric}).  The form for $g(x)$ compatible with
Eq.~(\ref{preMOND}) is
\be
	g(x) = \cases{ 
		x + c_1x^{2/3} &  Einstein ($x>x_c$)\cr
	 	\beta x^{2/3}\ln x + c_2x^{2/3} & MOND ($x<x_c$) , \cr
	}
\label{g}
\ee
for some constant parameters, $\beta$, $c_1$, and $c_2$, yielding
\be
a = \cases {-{1\over 2}{r_g\over  r^2} & $\vert a\vert > a_0$ \cr 
-{3\beta\over 2}{(r_gH_0)^{2/3}\over r} & $\vert a\vert < a_0$\ , \cr} \label{aMOND} \ee
where the critical MOND acceleration, $a_0$, is
\be
	a_0 = H_0\left[9\beta^2(r_gH_0)^{1/3}\right]\ .
\label{a0}
\ee
Observationally, we choose $\beta \approx 15$ so that for source
masses the size of large galaxies ($M\sim 10^{11}M_\odot$), the
critical acceleration is $a_0 \approx {1\over 6}H_0$, corresponding
to $x_c \approx 7\times 10^4$. A relationship
between $c_1$ and $c_2$ exists to ensure that $g(x)$ is continuous
across the the transition at $a = a_0$:
\be
	c_1 = c_2 + 3\beta\left[\ln (3\beta)-1\right]\ ,
\ee
and the remaining constant represent an arbitrary choice in zero-point
energy for the Newtonian potential.\footnote{
  Although the $c_1$ and $c_2$--terms do not affect the Newtonian
  acceleration, they do have a nontrivial affect on cosmology,
  simulating curvature-type terms in the Friedmann equation even
  though the cosmology is explicitly spatially-flat
  (see Eq.~(\ref{line-cosmo})).  Moreover, these terms have a
  effect on $g_{rr}$ (see Eq.~(\ref{metric})), manifesting in nontrivial,
  though immeasurably small, effects on gravitational lensing and
  the post-Newtonian parameter, $\gamma$.}

The form Eq.~(\ref{aMOND}) for the MOND gravitational acceleration is
slightly different than the form typically considered, i.e., one where
$a_0 = {1\over 6}H_0$ is a universal constant.  Compatibility with a
homogeneous cosmology compels us to choose a form where the critical
acceleration has a weak dependence on the source mass, $a_0\sim
M^{1/3}$.  This mass dependence is not a serious amendment to the MOND
paradigm, and indeed a well-motivated mass dependence may actually
benefit the modified-gravity scenario.  Galaxy clusters appear to
stray from the original MOND parametrization \cite{Sanders:2002pf},
such that uncomfortably large light-to-mass ratios are necessary to
bring objects of mass scales ${\cal O}(10^{14}M_\odot)$ into accord
with MOND.  In our prescription, Eq.~(\ref{aMOND}), objects more
massive than galaxies would exhibit effective dark matter halos somewhat
heavier than those predicted by traditional MOND, consistent with what
seems to be required by observations.

Let us summarize.  By requiring that the underlying fundamental
gravity theory that provides MOND Newtonian accelerations are
compatible with both homogeneous cosmologies and Birkhoff's law, we
may construct the function $g(x)$ found in Eq.~(\ref{g}) which
determines both the modified Friedmann equation, Eq.~(\ref{FRW}), and
the full Schwarzschild-like metric of a spherical mass source,
Eq.~(\ref{metric}), avoiding explicit reference to the details of the
fundamental theory.  With these two governing relationships, we may now
articulate a whole host of important properties of cosmological
interest.  For example, we may compute modifications of planetary
ephemeris, gravitational lensing, growth of density perturbations (both linear
and nonlinear), and the late-time integrated Sachs--Wolfe
(ISW) effect on the cosmic microwave background.  Details of these
calculations for arbitrary $g(x)$ are given in Ref.~\cite{Lue:2003ky}
Here we focus particularly on accommodating late-time acceleration into
MOND, and on the growth of fluctuations.

\section{MOND Cosmology}

Beginning with a function $g(x)$ given by Eq.~(\ref{g}) and
consistent with MOND Newtonian accelerations, Eq.~(\ref{aMOND}),
one can articulate a MOND cosmology.  The quantity $x$ that
appears in the modified Friedmann equation, Eq.~(\ref{FRW}), may
be interpreted as $\Omega_b(t)$ assuming that the matter content in the
universe is dominated by baryons. (At earlier times one would require
some MOND description of the self-gravity of radiation.)
One can then immediately associate $x$ with a redshift using the relationship
$x = (1+z)^3\Omega_b^{\rm today}$.  Let us investigate the
cosmology in stages, beginning with the Einstein, large--$x$,
stage.

\subsection{Early cosmology:  The CMB}

The transition from from Einstein to MOND takes place in galaxies
at $x\sim 7\times 10^4$, or correspondingly, taking
$\Omega_b^{\rm today} \approx 0.04$ \cite{WMAP},
cosmologically at a redshift $z\sim 120$. Thus, cosmology at redshifts
$z \gtrsim 120$ follows the ordinary GR Friedmann equation.  But if the matter content of the universe is solely baryonic, then matter--radiation equality
occurs at $z\sim 600$ whereas recombination still occurs at $z\sim
1100$, implying that recombination is {\em
before} radiation-matter equality rather than after it as in conventional
dark-matter cosmology.  
This observation corroborates the approach taken in prior work regarding
how MOND affects the cosmic microwave background (CMB)
\cite{McGaugh:1999cd,McGaugh:2000nn}.  The acoustic oscillations that
appear in the CMB anisotropy must, as expected, be driven in an almost
purely baryonic scenario.  This prior work claims that MOND not only
survives this drastic discrepancy from the standard cosmological model,
but that some ratios of CMB peak heights indeed favor MOND.  
Since our prescription applies strictly only during the matter-dominated epoch, 
it has nothing to contribute to the understanding of MOND physics at the epoch
of last scattering, although it can be used to make predictions about the
late-time integrated Sachs--Wolfe (ISW) effect.

\subsection{Late cosmology:  Recollapse, expansion and acceleration}

Looking at the form of Eq.~(\ref{g}), it is clear one cannot extend
MOND force-law to arbitrarily small $x$, or density.  Eventually,
the Hubble parameter, $H = \dot{a}/a$ vanishes and the
universe recollapses, regardless of the choice one makes for
$c_2$.  The result is intuitive if one imagines cosmology as
evolving classically on the Newtonian potential at some fixed energy dictated
by $c_2$.  The scale factor, $a$, is proportional to the position 
of a test particle on that potential.  The MOND part of the potential
is logarithmic, implying that every comoving trajectory eventually has
a turning point for any choice of initial energy (i.e., choice of $c_2$).

\begin{figure}
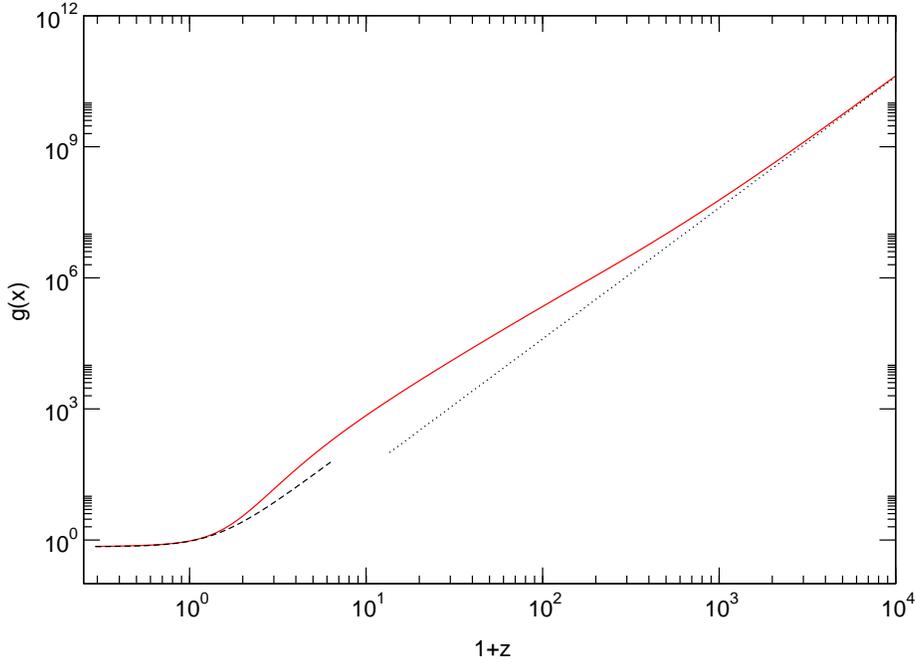
 \begin{center}\PSbox{g.eps
hscale=50 vscale=50 hoffset=-80 voffset=-25}{3in}{3.1in}\end{center}
\caption{
The function $g(x)$ versus $1+z$ where
$x = (1+z)^3\Omega_b^{\rm today}$.  The dotted line is $g(x)=x$,
the asymptotic Einstein behavior at high redshifts.  The dashed line
is $g(x) = \Omega_\Lambda + 6x$, which reproduces the late-time
expansion history of a standard dark-matter model.  The
interpolating $x^{2/3}\ln(1+x)$ dependence is the MOND regime.
}
\label{fig:g}
\end{figure}

Thus, there must be a sufficiently small $x$ where MOND
behavior ceases to predominate.  
We are guided here by the data which teaches us that:
\begin{enumerate}
\item Our Hubble expansion rate is currently 
$H \equiv H_0\simeq 70\ {\rm km/s}$.
(A value 20 or 30\% smaller than this would not change these arguments materially.)
\item We are currently undergoing acceleration in our cosmic
expansion \cite{Perlmutter:1998np,Riess:1998cb}
with $\ddot{a}/a \sim H_0^2$.
\item The expansion before $z\sim 1.7$ was decelerating\cite{Benitez:2002}.
\end{enumerate}
We can accommodate these considerations by modifying $g(x)$ 
of Eq.~(\ref{g}) in the following way:
\be
	g(x) = \cases{ 
	 x + 3\beta x^{2/3}\left[\ln(3\beta)-1\right] & $x \simge (3\beta)^3$  \quad\quad\quad\ \ (Einstein), \cr
	 \beta x^{2/3}\ln(1+x) &  $ 0.1 \simle x \simle (3\beta)^3$ \quad  (MOND), \cr
	 \Omega_\Lambda & $x \simle 0.1$  \quad\quad \quad\quad\ \ (``Dark Energy''), \cr
	}
\label{g-full}  
\ee
where $g(x)=\Omega_\Lambda \approx 0.7$ is equivalent to a cosmological
constant.  Figure~\ref{fig:g} shows these different regimes
and a possible smooth interpolation.  It is interesting that
such a simple modification may
be accommodated.  If $\beta$ were an order-of-magnitude
larger or smaller, one could not extend the MOND regime
all the way to the deceleration-acceleration transition and
still be able to maintain both $H \sim H_0$ as well as $\ddot{a}/a
\sim H_0^2$.

\subsection{Late cosmology:  Linear perturbation growth}

\begin{figure}
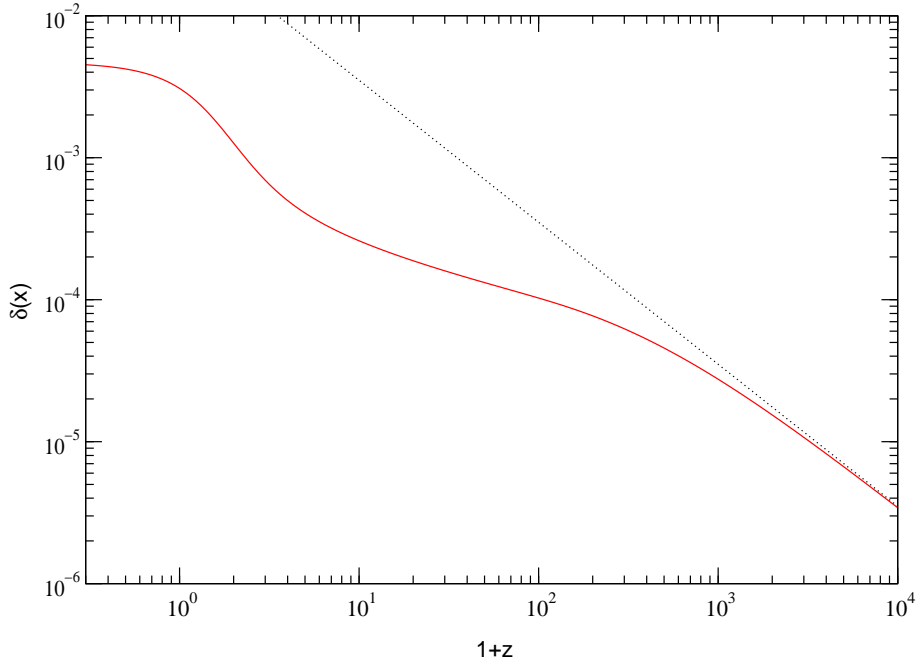
 \begin{center}\PSbox{delta.eps
hscale=50 vscale=50 hoffset=-80 voffset=-25}{3in}{3.1in}\end{center}
\caption{
Baryon density perturbation, $\delta(x)$ versus $1+z$ for the
function $g(x)$ depicted in Fig.~\ref{fig:g}, normalized to be
consistent with CMB anisotropy amplitude at recombination.  The
dotted line represents the growth in $\delta(x)$ if it were to follow
Einstein gravity.
}
\label{fig:delta}
\end{figure}

Equations~(\ref{FRW}) and~(\ref{g-full}) represent the full
modified Friedmann equation from matter-radiation equality
to the present time, including the onset of today's cosmic
acceleration.  We may now proceed to compute the growth
of linear perturbations in this cosmology.  Such a computation
is important because we require sufficient density perturbation
growth to seed the observed structure in the Universe.  The evolution of linear
density perturbations for the class of theories under consideration
take a simple closed form.  Take a uniform overdensity in a
localized spherical region such that
\be
	\rho(t) = \bar{\rho}(t)\left[1+\delta(t)\right]\ ,
\ee
where $\bar{\rho}$ is the background matter density that
follows cosmological evolution.  Parameterizing time-evolution
using $x = 8\pi G\bar{\rho}/3H_0^2$, the growing
perturbation mode $\delta(x)$ goes as
\cite{Lue:2003ky} (see also  \cite{Multamaki:2003vs}):
\be
     \delta(x) = {5A\over 6}
     g^{1/2}(x)\int_x^\infty {dy\over y^{1/3}g^{3/2}(y)}\ , 
\label{delta}
\ee
where $A$ is an overall normalization.  Because matter is
predominantly baryons, the normalization is fixed by requiring that 
perturbations are restricted to be $\delta \sim few \times 10^{-5}$ at recombination.
Figure~\ref{fig:delta} depicts the evolution of $\delta(x)$ for a
smooth interpolation of Eq.~(\ref{g-full}).

In pure matter-domination for Einstein FRW, $\delta = Ax^{-1/3}$,
or in other words, $\delta$ grows like the scale factor $a(t)$.  Even
if growth were as large as this, given the normalization required
at recombination, the growth of density perturbations would be
insufficient to account for the observed structure formation.  
Moreover, in each of the three
regimes in Eq.~(\ref{g-full}), growth is slower than that given
benchmark, $\delta < a(t)$.  In the early Einstein phase and during
the MOND phase, scale factor evolution looks as if it were
curvature-dominated.  Growth must take place before the end
of the MOND phase.

It may seem counterintuitive that growth is suppressed during the
MOND regime, given that in such a regime the self-gravitation of
overdensities should be enhanced.  But one must recall that the
same stronger gravity also drives a faster cosmology, which
in turn suppresses perturbation growth.  Ultimately, this latter
effect wins out.  This  poses a significant difficulty for
cosmological incarnations of MOND.

\subsection{Late cosmology:  Tuning the late-time potential}

There is a way to avoid this difficulty, but a
specially-tailored force law is required.  From galaxy
rotation curves, we require that MOND need only be valid up to
radii $r \sim 70\ {\rm kpc}$ for galaxy masses $M\sim 10^{11}M_\odot$
\cite{Sanders1996,deBlok1998}.  
This distance and mass scale corresponds to
$x \sim 600$ or, in MOND cosmology, to a redshift $z\sim 25$.  
If a recent observation
of a Gunn--Peterson trough \cite{GP} is a signal that galaxies
formed near $z\sim 6$ (or $x\sim 14$), then there is a narrow
window in $x$, between $14$ and $600$,
where little is known observationally and where
one can carefully manipulate $g(x)$ to achieve sufficient
growth in density perturbations to create galaxies, yet still
maintain the MOND paradigm.

To achieve the required growth,  somewhere in this range of $x$,
the function $g(x)$ must dip very close to zero
and then rises again above
${\cal O}(1)$ to accommodate SNIA constraints on
contemporary expansion history.  From Eq.~(\ref{delta}) one
sees that near a minimum where
$g(x) = g(x_0) + {1\over 2}g''(x_0)(x-x_0)^2$
\be
	\delta \sim {A\over x_0^{1/3}}{1\over\sqrt{g(x_0) g''(x_0)}}\ .
\ee	
When $g(x_0)$ is close to zero, arbitrarily large growth in $\delta(x)$
can occur.  The cosmology in this regime loiters, the expansion
almost stops and near this critical unstable point in the potential,
small variations in density amplify.  For a $M\sim 10^{11}M_\odot$
source mass, this fine-tuned dip in $g(x)$ corresponds to
Newtonian gravity becoming {\em repulsive} in a region
$r\sim 70\rightarrow 300\ {\rm kpc}$ to generate the large
perturbations and then becoming attractive again before
$r\sim 600\ {\rm kpc}$ to account for today's cosmology. 

But even this possible resolution is a tenuous one.  WMAP
observations of the CMB suggests that reionization starts as early as
$z\approx 20^{+10}_{-9}$ \cite{WMAP}, and that growth in
perturbations must occur before that redshift.  The window for
a possible excursion in $g(x)$ then becomes exceedingly small,
casting doubt that a MOND cosmology can viably create the
universe we see today.

\section{Concluding Remarks}

In this paper, we provided a self-consistent framework where
we could assess which cosmologies were compatible with
Modified Newtonian Dynamics (MOND), exploiting techniques
developed in prior work \cite{Lue:2003ky}.  
Our starting point was the MOND paradigm that the contents 
of the universe are what we see, and that these alone drive the
dynamics of the cosmic expansion.
We find that in order for MOND to exhibit homogeneous cosmologies, 
the critical MOND acceleration $a_0$, which separates Einstein
behavior from a modified Newtonian force, must have a slight
source-mass dependence ($a_0\sim M^{1/3})$.

With that mild amendment, we found that MOND cosmologies are
naturally compatible with observed late-time expansion histories
and late-time cosmic acceleration. However, those natural
cosmologies cannot produce enough growth  of density perturbations
to account for structure formation.
Two effects contribute:
first, because matter is almost exclusively baryonic, matter
perturbations are constrained to be ${\cal O}(10^{-5})$ at
recombination;  furthermore, during those redshifts where MOND
behavior dominates, growth of perturbations is suppressed,
rather than enhanced as one might have expected.  
Although gravity is stronger than usual
in this regime, the potential enhanced self-gravitation of density
fluctuations is beaten by the faster expansion
at a given redshift required 
if the Hubble parameter is to have its observed value today 
despite the stronger gravity  that MOND predicts.

One may circumvent this difficulty by envisioning a loitering
phase, arising from a drastic weakening of gravity at a selected value of $x$,
and chosen to correspond to part of redshift history where
little is known ($z \sim 6\rightarrow 25$) -- as the   
the cosmic expansion stalls, the self-gravitation of perturbations proceeds
unhindered.  These cosmologies are contrived and fine-tuned, 
and may even be marginally
ruled out by evidence of early ($z \sim 20$) reionization.  Such
machinations cast doubt on the possibility that MOND cosmology
can viably lead to the universe we observe today.

\acknowledgments

The authors would like to thank S.~McGaugh, R.~Scoccimarro and J.-P. Uzan
for helpful communications and insights.  We are also grateful to
the CERN Theory Division for their hospitality.  This work is
sponsored by DOE Grant DEFG0295ER40898, the CWRU Office of the Provost,
and CERN.

\end{document}